\documentclass[11pt,twoside,twocolumn,a4paper]{article}

\usepackage{a4wide,amsfonts,amsmath,amssymb,bm,abstract,balance,natbib,graphicx,subfigure}

\title{\textbf{Attractive interaction between ions inside a quantum plasma structure}}

\author{Maxim Dvornikov$^{1,2,3,4}$
\\
\small{$^{1}$Physics Faculty, National Research Tomsk State University, 36 Lenin Ave., 634050 Tomsk, Russia} \\
\small{$^{2}$Institute of Physics, University of S\~{a}o Paulo, CP 66318, CEP 05314-970 S\~{a}o Paulo, SP, Brazil} \\
\small{$^{3}$Pushkov Institute of Terrestrial Magnetism, Ionosphere and Radiowave Propagation (IZMIRAN),} \\
\small{142190 Troitsk, Moscow, Russia} \\
\small{$^{4}$Nonlinear Physics Centre, RSPE, Australian National University, 2601 Canberra, ACT, Australia} \\
\small{E-mail: maxim.dvornikov@usp.br}}

\date{}

\begin{document}


\twocolumn[\maketitle
\begin{onecolabstract}
We construct the model of a quantum spherically symmetric plasma structure
based on radial oscillations of ions. We suppose that ions are involved
in ion-acoustic plasma oscillations. We find the exact solution of the Schr\"{o}dinger
equation for an ion moving in the self-consistent oscillatory potential
of an ion-acoustic wave. The system of ions is secondly quantized
and its ground state is constructed. Then we consider the interaction
between ions by the exchange of an acoustic wave. It is shown that
this interaction can be attractive. We describe the formation of pairs
of ions inside a plasma structure and demonstrate that such a plasmoid
can exist in a dense astrophysical medium corresponding to the outer core of a neutron star.
\end{onecolabstract}]


\section{Introduction}

The Coulomb interaction is undoubtedly dominant for charged particles
in plasma. For instance, various oscillatory processes, like Langmuir
waves, are driven by the Coulomb interaction. However, in the last
decades there is a growing interest for various effective interactions
which can arise between charged particles in plasmas.

First we mention that \citet{NamVlaShu95} studied the wakefield interaction between particles of the
same polarity moving in plasma. Under certain
conditions the wakefield interaction can be attractive. \citet{MorIvl09} found that this kind
of interaction can result in the formation of complex structures in
dusty plasmas. Recently \citet{Car12} observed the attractive wakefield
interaction in the ions system in a laboratory experiment.

Besides the wakefield interaction which arises mainly in classical plasmas,
\citet{Bon08} showed that an effective interaction between charged particles in dense
strongly correlated plasmas is responsible for the formation
of Coulomb crystals. Note that Coulomb crystals can
exist in both classical and quantum plasmas.
\citet{GriAda79} observed Coulomb crystals in liquid helium, \citet{BirKasWal92} in a system of trapped
laser cooled ions, and \citet{ChuL94}
in dusty plasmas.
\citet{Bai09} suggested that magnetized Coulomb crystals play an important role in the evolution of neutron stars.

Another type of the effective interaction between charged particles
in plasma is based on the exchange of a virtual acoustic wave.
\citet{VlaYak78} proposed that this effect can explain the stability
of atmospheric plasmoids. Recently \citet{Dvo13}
studied analogous effective interaction
to describe the pairing of ions performing
spherically symmetric quantum oscillations. The weakness of the description
of quantum plasmoids undertaken by \citet{Dvo13} consists in the assumption
that ions participate in Langmuir oscillations which is valid only
in the short waves limit.

In the present work we continue the study of spatially localized
quantum plasma structures based on radial oscillations of charged
particles. To describe the motion of charged particles we do not use
quantum hydrodynamics since in that approach quantum effects are
accounted for perturbatively. In Sec.~\ref{sec:QUANTSTAT}
we find the exact solution of the Schr\"{o}dinger equation describing
the motion of an ion in the self-consistent potential of an ion-acoustic
wave. Using the second quantization method we construct the ground
state of a plasmoid corresponding to the oscillatory motion of ions. In Sec.~\ref{sec:EFFINT}
we consider the effective interaction between oscillating ions by
the exchange of an acoustic wave. In particular, we show that this
interaction can be attractive. In Sec.~\ref{sec:APPL} we discuss
the pairing of ions due to this attractive interaction and demonstrate
that a plasma structure, where bound states of ions are formed, can
well exist in a dense matter of the outer core of a neutron star
(NS).

Finally, in Sec.~\ref{sec:CONCL}, our
results are briefly summarized.
The mathematical details of the analysis of ions wave functions
in the coordinate representation are given in Appendix~\ref{sec:WFCOORD}.

\section{Quantum states of ions in a spherical plasmoid\label{sec:QUANTSTAT}}

In this section we develop the model of a quantum plasmoid based on
radial oscillations of ions in plasma. We find the exact solution
of the Schr\"{o}dinger equation describing a spherically symmetric motion
of ions participating in ion-acoustic oscillations. Then this system
is secondly quantized and its ground state is constructed. We formulate
the criterion of the stability of quantum oscillations and consider
the description of the plasmoid in the short waves limit.

One of the most popular approaches to account for quantum effects in plasmas is based on the quantum hydrodynamics, recently reviewed by~\citet{Haa11}. In this method, the classical hydrodynamic equations are used to study the plasma dynamics. Quantum effects are accounted for by adding the quantum pressure term. Thus, by construction, within the quantum hydrodynamics, plasmas are mainly classical and quantum effects are treated perturbatively. It is a reasonable approximation for a high temperature plasma.

On the contrary, if one studies a strongly correlated system of charged particles, e.g., a degenerate plasma (as an example, see Sec.~\ref{sec:EFFINT} below) the dynamics of particles in such a medium should obey mainly quantum mechanical laws. \citet{BonPehSch13} showed that the methods of the quantum hydrodynamics are not applicable in this situation.


In our work we consider the model of a quantum plasma structure based on
spherically symmetric oscillations of charged particles.
As a rule, plasma oscillations are dominated by the motion an electron
component of plasma. It happens since an electron is a very light
particle compared to an ion. Despite the amplitude and the frequency
of electrons oscillations are much higher than that of ions, in realistic
situation we cannot neglect the ions motion. Typically ions have lower
temperature. Thus, quantum effects can be more pronounced for them.
Note that, in Sec.~\ref{sec:APPL}, we shall study the effective interaction
of ions in a very dense matter, with the temperature being much less than the chemical potential. It justifies the application of the
quantum approach for the description of the ions motion.

Let us suppose that ions in plasma oscillate with the frequency $\omega$.
If we consider an ion as a test particle, the stationary Schr\"{o}dinger equation for the ion wave function
$\psi$ has the form,
\begin{equation}\label{eq:Scheqgen}
  E\psi =
  \hat{H}\psi,\quad\hat{H} =
  \frac{\hat{\mathbf{p}}^{2}}{2m_{i}} +
  \frac{m_{i}\omega^{2}\hat{\mathbf{r}}^{2}}{2},
\end{equation}
where $\hat{\mathbf{p}}$ and $\hat{\mathbf{r}}$ are the operators
of the momentum and the coordinate, $E$ is the ion energy, and $m_{i}$
is its mass.

\citet{Dvo13} studied the quantization of the motion of
charged particles participating in Langmuir oscillations, which is
a rough model in case we deal with ions. Let us suppose that ions
in plasma are involved in collective oscillations corresponding to
ion-acoustic waves. Thus the frequency $\omega$ in Eq.~(\ref{eq:Scheqgen})
obeys the dispersion relation~\citep{LifPit10},
\begin{equation}\label{eq:disprel}
  \omega^{2} =
  \omega_{i}^{2}\frac{\lambda_{e}^{2}k^{2}}{1+\lambda_{e}^{2}k^{2}},
\end{equation}
where $\omega_{i}$ is the Langmuir frequency
for ions,
$\lambda_{e}$
is the Debye
length for electrons, and $k=|\mathbf{k}|$
is the absolute value of the wave vector. If we consider a classical electron-ion plasma, then $\omega_{i}=\sqrt{4\pi e^{2}n_i^{(0)}/m_{i}}$ and $\lambda_{e}=\sqrt{T_{e}/4\pi e^{2}n_e^{(0)}}$, where $T_{e}$ is the electron temperature and $n_{i,e}^{(0)}$ are the unperturbed densities of ions and electrons.

In Sec.~\ref{sec:APPL} we shall study quantum plamoids in a degenerate plasma consisting of ultrarelativistic electrons and nonrelativistic ions (protons). Using the results of~\citet{BraSeg93} one gets that the dispersion relation for ion-acoustic waves in this case has the same form as in Eq.~\eqref{eq:disprel}. However, $\lambda_{e} =\tfrac{c}{3\omega_e}$, where $c$ is the speed of light. The Langmuir frequency
for electrons $\omega_e$ as well as $\omega_i$ have the form~\citep{BraSeg93},
\begin{align}\label{eq:Langfrdegpl}
  \omega_e^2 = & \frac{4\alpha_\mathrm{em}}{3\pi}[3\pi^2 n_e^{(0)} c^3]^{2/3},
  \notag
  \\
  \omega_i^2 = & \frac{4\alpha_\mathrm{em}}{3\pi}
  \frac{3\pi^2 n_i^{(0)}}{m_i} \hbar c,
\end{align}
where $\alpha_\mathrm{em} \approx 7.3 \times 10^{-3}$ is the fine structure constant.

We shall choose the momentum representation in Eq.~(\ref{eq:Scheqgen}),
i.e. $\hat{\mathbf{p}}=\mathbf{p}=\hbar\mathbf{k}$ and $\hat{\mathbf{r}}=\mathrm{i}\hbar\nabla_{\mathbf{p}}$,
where $\nabla_{\mathbf{p}}=\tfrac{\partial}{\partial\mathbf{p}}$.
Using Eq.~(\ref{eq:disprel}), we can rewrite Eq.~(\ref{eq:Scheqgen})
as
\begin{equation}\label{eq:Scheqp}
  \bigg(
    \frac{p^{2}}{2m_{i}} -
    \frac{m_{i}\omega_{i}^{2}}{2}
    \frac{\lambda_{e}^{2}p^{2}}{\hbar^{2}+\lambda_{e}^{2}p^{2}}
    \hbar^{2}\nabla_{\mathbf{p}}^{2}
  \bigg)
  \psi = E\psi,
\end{equation}
where $\psi=\psi(\mathbf{p})$. The coefficients in Eq.~(\ref{eq:Scheqp})
depend on the absolute value of the momentum $p=|\mathbf{p}|$. Thus
we can look for the solution of Eq.~(\ref{eq:Scheqp}) as $\psi(\mathbf{p})=R(p)Y_{l\mathrm{m}}(\theta,\phi)$,
where $R$ is the radial part of the wave function, $Y_{l\mathrm{m}}(\theta,\phi)$
is the spherical harmonic corresponding to the orbital and magnetic
quantum numbers: $l=0,1,2,\dotsc$ and $\mathrm{m}=0,\pm1,\dots,\pm l$,
$\theta$ and $\phi$ are the spherical angles fixing the momentum
direction. Note that Eq.~\eqref{eq:Scheqp} is analogous to that studied by~\citet{Alh02,Sch07}.

Let us introduce the new variable $\rho=p^{2}/m_{i}\omega_{i}\hbar$
and the new unknown function $u$, as $R=e^{-\rho/2}\rho^{l'/2}u(\rho)$,
where
\begin{equation}\label{eq:l'l}
  l'=\frac{1}{2}
  \left\{
    \left[
      (2l+1)^{2}-\frac{8E}{m_{i}\omega_{i}^{2}\lambda_{e}^{2}}
    \right]^{1/2}-1
  \right\},
\end{equation}
is the effective orbital quantum number. Note that $-1/2\leq l'\leq l$.
In these new variables Eq.~(\ref{eq:Scheqp}) takes the form,
\begin{multline}\label{eq:Schequ-1}
  \rho\frac{\mathrm{d}^{2}u}{\mathrm{d}\rho^{2}} +
  \left(
    \frac{3}{2}+l'-\rho
  \right)
  \frac{\mathrm{d}u}{\mathrm{d}\rho} 
  \\
  +
  \left(
    \frac{E'}{2\hbar\omega_{0}}-\frac{3}{4}-\frac{l'}{2}
  \right)
  u=0,
\end{multline}
where $E'=E-\tfrac{\hbar^{2}}{2m_{i}\lambda_{e}^{2}}$ is the effective
energy.

The solution of Eq.~(\ref{eq:Schequ-1}) can be expressed using Eq.~(9.216) on page~1024 in \citet{GraRyz07},
\begin{equation}\label{eq:solSchequ-1}
  u = {}_{1}F_{1}
  \left(
    -\frac{E'}{2\hbar\omega_{0}}+\frac{l'}{2}+\frac{3}{4};l'+\frac{3}{2};\rho
  \right),
\end{equation}
where $_{1}F_{1}(a;b;z)$ is the Kummer's confluent hypergeometric
function. The hypergeometric function in Eq.~(\ref{eq:solSchequ-1})
should be finite at $\rho\to\infty$. Thus the energy of an ion should
satisfy the relation,
\begin{align}\label{eq:enspectrl-1}
  E_{nl} = & \hbar\omega_{i}
  \bigg\{
    2n+1-\frac{\hbar}{2m_{i}\omega_{i}\lambda_{e}^{2}}
    \notag
    \\
    & +
    \bigg[
      \left(
        l+\frac{1}{2}
      \right)^{2} -
      4
      \left(
        n+\frac{1}{2}
      \right)
      \notag
      \\
      & \times
      \frac{\hbar}{m_{i}\omega_{i}\lambda_{e}^{2}}
    \bigg]^{1/2}
  \bigg\},
\end{align}
where $n=0,1,2,\dotsc$ is the radial quantum number. It should be
noted that, at big $n$, ions oscillations become unstable since the
energy in Eq.~(\ref{eq:enspectrl-1}) acquires the imaginary part.
Thus we should impose a restriction,
\begin{equation}\label{eq:ncr}
  n<n_{\mathrm{cr}} =
  \left(
    l+\frac{1}{2}
  \right)^{2}
  \frac{m_{i}\omega_{i}\lambda_{e}^{2}}{4\hbar} -
  \frac{1}{2},
\end{equation}
to guarantee the stability of oscillations.

In the following we shall study spherically symmetric plasma oscillations.
Thus we should consider wave functions independent of $\theta$ and
$\phi$. This case corresponds to $l=0$. Using Eq.~(\ref{eq:solSchequ-1})
one finds the properly normalized total wave function, which also
includes the spin variables, in the following form:
\begin{align}\label{eq:wfsphsym}
  \psi_{n\sigma}(p)= & 2\pi
  \left[
    \frac{n!}{\Gamma(l'+n+3/2)}
  \right]^{1/2}
  \notag
  \\
  & \times
  \left(
    \frac{\hbar}{m_{i}\omega_{i}}
  \right)^{3/4}
  \left(
    \frac{p^{2}}{m_{i}\omega_{i}\hbar}
  \right)^{l'/2}
  \nonumber
  \\
  & \times
  \exp
  \left[
    -\frac{p^{2}}{2m_{i}\omega_{i}\hbar}
  \right]
  \notag
  \\
  & \times
  L_{n}^{l'+1/2}
  \left(
    \frac{p^{2}}{m_{i}\omega_{i}\hbar}
  \right)\chi_{\sigma},
\end{align}
where $L_{n}^{\alpha}(z)$ is the associated Laguerre polynomial,
$\Gamma(z)$ is the Euler gamma function, $\chi_{\sigma}$ is the
spin wave function, and $\sigma$ is the spin variable.

Note that both Eqs.~\eqref{eq:Scheqgen} and~\eqref{eq:Scheqp} correspond to a spinless ion. The spin dependence was introduced as a direct product of the coordinate wave function $\psi_{n}(p)$ and the spin wave function $\chi_{\sigma}$. Thus we do not consider here effects related to the spin-orbit interaction like in the Dirac theory.

The new expression
for $l'$ has the form,
\begin{equation}\label{eq:l'sphsym}
  l'=\frac{1}{2}
  \left\{
    \left[
      1-\frac{8E}{m_{i}\omega_{i}^{2}\lambda_{e}^{2}}
    \right]^{1/2} -
    1
  \right\}.
\end{equation}
Note that now $-1/2\leq l'\leq0$. The energy levels corresponding
to the states described by $\psi_{n\sigma}$ in Eq.~(\ref{eq:wfsphsym})
have the form,
\begin{align}\label{eq:enspectsphsym}
  E_{n} = & \hbar\omega_{i}
  \bigg\{
    2n+1-\frac{\hbar}{2m_{i}\omega_{i}\lambda_{e}^{2}}
    \notag
    \\
    & +
    \bigg[
      \frac{1}{4} - 4
      \left(
        n+\frac{1}{2}
      \right)
      \notag
      \\
      & \times
      \frac{\hbar}{m_{i}\omega_{i}\lambda_{e}^{2}}
    \bigg]^{1/2}
  \bigg\}.
\end{align}
Note that Eqs.~\eqref{eq:l'sphsym} and~\eqref{eq:enspectsphsym} can be obtained directly from Eqs.~\eqref{eq:l'l} and~(\ref{eq:enspectrl-1}) and correspond to the case $l=0$.

We shall assume that ions are singly ionized. Thus they should be
fermions. Indeed, a typical neutral atom has an integer spin. Thus,
if we remove one electron, an ion becomes a fermion. For simplicity
we shall assume that ions have the lowest possible spin, i.e. they
are $1/2$-spin particles. In Sec.~\ref{sec:APPL} we shall study a degenerate electron-proton plasma in the NS outer core. In this case an ion (a proton) certainly has the spin equal to $1/2$. It means that $\sigma=\pm1$ in Eq.~(\ref{eq:wfsphsym}).

Now the ground state of the system can be constructed. Suppose that
we have $N$ ions performing ion-acoustic oscillations.
On the basis of Eq.~(\ref{eq:wfsphsym}) we can introduce the operator
valued wave function,
\begin{equation}\label{eq:wfsq}
  \hat{\psi} = \sum_{n\sigma}\psi_{n\sigma}\hat{a}_{n\sigma},
\end{equation}
and the analogous expression for $\hat{\psi}^{\dagger}$, which contains
$\hat{a}_{n\sigma}^{\dagger}$. Here $\hat{a}_{n\sigma}^{\dagger}$
and $\hat{a}_{n\sigma}$ are the creation and annihilation operators
of the states corresponding to ions oscillations. These operators obey the usual
anticommutation relation, $\left[ \hat{a}_{n\sigma}, \hat{a}_{n'\sigma'}^{\dagger} \right]_{+} = \delta_{nn'}\delta_{\sigma\sigma'}$,
with other anticommutators being equal to zero. In this situation  each energy state can be
occupied by no more than two particles. Thus, we do not contradict the Pauli principle.

The constructed ground state corresponds to the collective motion of ions in a spherically symmetric plasma structure since the occupied energy levels are associated with particles involved in ion-acoustic oscillations. It should be noted that in our model we consider ions as test particles oscillating
with the same frequency $\omega$ corresponding to an ion-acoustic wave; cf. the effective potential in Eq.~\eqref{eq:Scheqgen}. This approximation is valid if we neglect temperature
effects~\citep{Daw59}.

From the formal point of view, we can choose any complete set of basis wave functions to build a ground state. However, for the description of the motion of ions which are involved
in spherically symmetric oscillations, the ground state constructed in our work is preferable since it accounts for the dynamical features of
the system and its geometric symmetry. It is not the case for a ground state based on plane waves, $\sim \exp(\mathrm{i}\mathbf{kr})$,
which are typically chosen to describe the evolution of test particles in quantum plasmas. In other words, the ground state constructed takes into account the electromagnetic interactions between charged particles in plasma in all orders in a perturbative expansion. Indeed, plasma oscillations in our system are driven by electromagnetic interactions and we use the basis wave functions corresponding to a 3D harmonic oscillator.

The energy of the ground state can be found as $E_{0}=\left\langle \hat{H}_{\mathrm{ion}}\right\rangle $,
where
\begin{equation}\label{eq:Hion}
  \hat{H}_{\mathrm{ion}} =
  \sum_{n\sigma}E_{n}\hat{a}_{n\sigma}^{\dagger}\hat{a}_{n\sigma},
\end{equation}
is the ground state Hamiltonian of ions and the values of the energy
are given in Eq.~(\ref{eq:enspectsphsym}). Note that, since we have
a great but limited number of ions involved in oscillations,
\begin{equation}
  E_{0}\approx2\sum_{n<n_{\mathrm{F}}}E_{n},
\end{equation}
where $n_{\mathrm{F}}$ is the number of occupied energy states. We
shall call it the \emph{Fermi number}.

We can define the ``size'' of a spherically symmetric plasmoid in
the momentum space as the last maximum of the function $|\psi_{n_{\mathrm{F}}}(p)|^{2}$.
\citet{Blo64} showed that this maximum is approximately achieved at the classical turn point,
$p_{\mathrm{max}}^{2}=2m_{i}E_{n}$. Supposing that $n_{\mathrm{F}}\sim n_{\mathrm{cr}}$
and using Eq.~(\ref{eq:ncr}) we get that $p_{\mathrm{max}}\lesssim m_{i}\omega_{i}\lambda_{e}$.
If we study a nondegenerate electron-ion plasma and employ the quasiclassical limit, we can take that $p_{\mathrm{max}}\sim m_{i}v_{i}$,
where $v_{i}$ is the typical ion velocity. Thus we get that the constraint
in Eq.~(\ref{eq:ncr}) is equivalent to $L_{i}\sim v_{i}/\omega_{i}\lesssim\lambda_{e}$,
where $L_{i}$ is the typical particle displacement in ion oscillations.
It is the well known condition for the existence of ion-acoustic waves in a nondegenerate plasma~\citep{LifPit10}.

Note that for the sufficiently short waves the frequency of ion oscillations
is constant $\omega=\omega_{i}$; cf. Eq.~(\ref{eq:disprel}). In
the quantum description this limit is equivalent to $\lambda_{e}\gg\sqrt{\hbar/m_{i}\omega_{i}}$.
In this case the energy spectrum in Eq.~(\ref{eq:enspectsphsym})
coincides with that corresponding to Langmuir oscillations of ions obtained by~\citet{Dvo13}.
Besides the coincidence of the spectra for big $\lambda_{e}$, the
wave function in Eq.~(\ref{eq:wfsphsym}) is also consistent with
the result of ~\citet{Dvo13}. Indeed, taking into account that
$l'\to0$ at $\lambda_{e}\gg\sqrt{\hbar/m_{i}\omega_{i}}$, we get
the following expression for the Fourier transform of the wave function
in Eq.~(\ref{eq:wfsphsym}):
\begin{align}\label{eq:wfasympt}
  \psi_{n}(\mathbf{r})= & \int\frac{\mathrm{d}^{3}\mathbf{p}}{(2\pi\hbar)^{3}}
  e^{\mathrm{i}\frac{(\mathbf{pr})}{\hbar}}\psi_{n}(p)
  \notag
  \\
  & =
  \frac{1}{\sqrt{4\pi(2n+1)!}}\frac{1}{2^{n}}
  \left(
    \frac{m_{i}^{3}\omega_{i}^{3}}{\pi\hbar^{3}}
  \right)^{1/4}
  \nonumber
  \\
  & \times
  \frac{r_{0}}{r}
  \exp
  \left(
    -\frac{r^{2}}{2r_{0}^{2}}
  \right)
  \notag
  \\
  & \times
  H_{2n+1}
  \left(
    \frac{r}{r_{0}}
  \right),
\end{align}
where $H_{n}(z)$ is the Hermite polynomial and $r_{0}=\sqrt{\hbar/m_{i}\omega_{i}}$.
One can notice that in Eq.~(\ref{eq:wfasympt}) we reproduce the
result of~\citet{Dvo13} up to the sign factor. The details of
the derivation of Eq.~(\ref{eq:wfasympt}) are provided in Appendix~\ref{sec:WFCOORD};
cf. Eq.~(\ref{eq:wfexplr}). Note that in Eq.~\eqref{eq:wfasympt} we consider only the coordinate part of the wave function omitting $\chi_\sigma$.

\section{The effective interaction of ions\label{sec:EFFINT}}

In this section we study the effective interaction of ions in a spherically
symmetric plasmoid owing to the exchange of virtual acoustic waves.
It is demonstrated that this interaction can be attractive.

We consider the situation when the plasma temperature is not so high.
It means that a neutral component can be present. In Sec.~\ref{sec:APPL}
we will study plasma structures in the NS outer core, where a neutral component,
consisting of neutrons, is always present. In this case rapidly oscillating
ions will collide with neutral particles and generate acoustic waves.
If an acoustic wave is coherently absorbed inside the system, it will
result in the effective interaction between charged particles. One
can expect that this effective interaction is more efficient for ions
rather than for electrons.

We shall study oscillating ions in highly excited states with $n\gg1$.
Using Eqs.~(\ref{eq:wfasyptl0}) and~(\ref{eq:wfasymptl1/2}), we
get that the asymptotic form of the wave functions of such ions in the
coordinate representation coincides with that in Eq.~(\ref{eq:wfasympt}).
Note that the analogous form of the wave function for a charged particle performing radial oscillations in plasma was derived by~\citet{DvoDvo06}. One can prove it using Eq.~(8.955.2) on page~997 in~\citet{GraRyz07}. Therefore,
for $n \gg 1$, the effect of the spatial dispersion of ion-acoustic
waves does not significantly contribute to the form of the ions wave
functions.


If $\delta n(\mathbf{r})$ is the perturbation of the neutral particles density because of the acoustic wave propagation, then the energy of the interaction of an ion with the field of acoustic waves is
\begin{equation}\label{eq:Uionneut}
  U_\mathrm{int}(\mathbf{r}) =
  \int
  \mathrm{d}^{3}\mathbf{r}'
  K(\mathbf{r} - \mathbf{r}')
  \delta n(\mathbf{r}'),
\end{equation}
where $K(\mathbf{r} - \mathbf{r}')$ is the potential of the interaction between an ion and a neutral particle which are at the points $\mathbf{r}$ and $\mathbf{r}'$ respectively. Using Eq.~\eqref{eq:Uionneut} and the results of~\citet{VlaYak78}, we derive the secondly quantized Hamiltonian describing the interaction between ions and neutral particles as,
\begin{align}\label{eq:Helphrsq}
  \hat{H}_{\mathrm{int}} = &
  \int
  \mathrm{d}^{3}\mathbf{r}
  \hat{\psi}^{\dagger}(\mathbf{r})
  \hat{\psi}(\mathbf{r})
  U_\mathrm{int}(\mathbf{r}) 
  \notag
  \\
  & =
  \int
  \mathrm{d}^{3}\mathbf{r}
  \mathrm{d}^{3}\mathbf{r}'
  K(\mathbf{r} - \mathbf{r}')
  \notag
  \\
  & \times
  \hat{\psi}^{\dagger}(\mathbf{r})
  \hat{\psi}(\mathbf{r})
  \delta \hat{n}(\mathbf{r}'),
\end{align}
where the expression
for $\hat{\psi}$ is given in Eq.~(\ref{eq:wfsq}). Note that in Eq.~\eqref{eq:Helphrsq} we replaced $\delta n(\mathbf{r})$ with its secondly quantized analog $\delta \hat{n}(\mathbf{r})$; cf. Eq.~\eqref{eq:acfquant} below.

In the following we shall use the approximation of the contact interaction. It means that $K(\mathbf{r}-\mathbf{r}')=K_{0}\delta^{3}(\mathbf{r}-\mathbf{r}')$,
where $K_{0}$ is the phenomenological constant characterizing the
strength of the interaction. This approximation corresponds to a very short range interaction between ions and neutral particles. In Sec.~\ref{sec:APPL} we shall study a particular case of the proton-neutron interaction driven by nuclear forces which are short range. Thus the contact interaction approximation is justified.

The expression for $\delta \hat{n}(\mathbf{r})$ was obtained by~\citet{Dvo13},
\begin{align}\label{eq:acfquant}
  \delta \hat{n}(r) = &
  \int\frac{\mathrm{d}^{3}\mathbf{k}}{(2\pi)^{3/2}}
  \left(
    \frac{\hbar n_{n}^{(0)}}{2m_{n}\omega_{k}}
  \right)^{1/2}
  \notag
  \\
  & \times
  kf_{k}
  \left(
    \hat{b}_{k}+\hat{b}_{k}^{\dagger}
  \right),
\end{align}
where $f_{k}=\tfrac{\sin(kr)}{kr}$ is the spherically symmetric solution of
the wave equation for acoustic waves, $\hat{b}_{k}^{\dagger}$ and
$\hat{b}_{k}$ are the creation and annihilation operators for phonons,
$\omega_{k}$ is the frequency of acoustic oscillations, $n_{n}^{(0)}$
is the unperturbed density of neutral particles, and $m_{n}$ is the
mass of a neutral particle.

Using Eq.~\eqref{eq:acfquant}, we can rewrite Eq.~(\ref{eq:Helphrsq}) in the form,
\begin{align}\label{eq:Helphsq}
  \hat{H}_{\mathrm{int}} = &
  \int\frac{\mathrm{d}^{3}\mathbf{k}}{(2\pi)^{3/2}}
  \sum_{ns\sigma}
  D_{ns}(k)
  \notag
  \\
  & \times
  \hat{a}_{n\sigma}^{\dagger}\hat{a}_{s\sigma}
  \left(
    \hat{b}_{k}+\hat{b}_{k}^{\dagger}
  \right),
\end{align}
where $D_{ns}$ is the matrix element of this interaction. If we study
the effective interaction between ions occupying the same energy level,
we have for $D_{n}\equiv D_{nn}$,
\begin{align}\label{eq:Dnmapr}
  D_{n}(k)= & K_{0}k
  \left(
    \frac{n_{n}^{(0)}m_{i}\omega_{i}}{2m_{n}\omega_{k}}
  \right)^{1/2}
  \notag
  \\
  & \times
  \int\mathrm{d}^{3}\mathbf{r}
  \psi_{n}^2(r)
  f_{k}(r)
  \nonumber
  \\
  & \approx
  \frac{K_{0}}{4\sqrt{n}}
  \left(
    \frac{n_{n}^{(0)}m_{i}\omega_{i}}{2m_{n}\omega_{k}}
  \right)^{1/2}
  \notag
  \\
  & \times
  [1-\mathrm{sgn}(\xi-4\sqrt{n})],
\end{align}
where $\xi=k\sqrt{\hbar/m\omega_{i}}$ and $\text{sgn}(z)$ is the
sign function.

After the standard elimination of the acoustic degrees of freedom
with help of the canonical transformation,
\begin{align}
  & \hat{H}_{\mathrm{int}} \to
  \exp
  \left(
    -\hat{S}
  \right)
  \hat{H}_{\mathrm{int}} \exp
  \left(
    \hat{S}
  \right),
  \\
  & \hat{S} = \int\frac{\mathrm{d}^{3}\mathbf{k}}{(2\pi)^{3/2}}
  \sum_{ns\sigma}
  D_{ns}(k) \hat{a}_{n\sigma}^{\dagger} \hat{a}_{s\sigma}
  \notag
  \\
  \notag
  & \times
  \left(
    \frac{\hat{b}_{k}}{E_{s}-E_{n}-\hbar\omega_{k}}
    +
    \frac{\hat{b}_{k}^{\dagger}}{E_{s}-E_{n}+\hbar\omega_{k}}
  \right),
\end{align}
the total Hamiltonian, which includes $\hat{H}_{\mathrm{ion}}$ given
in Eq.~(\ref{eq:Hion}), takes the form,
\begin{align}\label{eq:Htotaps}
  \hat{H} = &
  \sum_{n\sigma}
  E_{n}
  \hat{a}_{n\sigma}^{\dagger}
  \hat{a}_{n\sigma}
  \notag
  \\
  & -
  \sum_{nn'\sigma}
  F_{nn'}
  \hat{a}_{n\sigma}^{\dagger}
  \hat{a}_{n',-\sigma}^{\dagger}
  \hat{a}_{n',-\sigma}
  \hat{a}_{n\sigma}.
\end{align}
To derive Eq.~(\ref{eq:Htotaps}) it is important to assume that
two interacting ions are at the same energy level. In Eq.~(\ref{eq:Htotaps})
we also account for the fact that these ions must have oppositely
directed spins because of the Pauli principle.

The amplitude of the effective interaction $F_{nn'}$ in Eq.~(\ref{eq:Htotaps})
has the form,
\begin{align}\label{eq:amplint}
  F_{nn'} = &
  \frac{K_{0}^{2}n_{n}^{(0)}m_{i}\omega_{i}}{32m_{n}\hbar\sqrt{nn'}}  \int\frac{\mathrm{d}^{3}\mathbf{k}}{(2\pi)^{3}}
  \frac{1}{\omega_{k}^{2}}
  \notag
  \\
  & \times
  [1-\mathrm{sgn}(\xi-4\sqrt{n})]
  \notag
  \\
  & \times  
  [1-\mathrm{sgn}(\xi-4\sqrt{n'})].
\end{align}
Now we should take into account the fact that ions and neutral particles
have practically the same mass. Thus the energy transfer in their
collisions occurs rather effectively. Therefore we should take that
the frequency of virtual acoustic waves in Eq.~(\ref{eq:amplint})
is close to the frequency of ion-acoustic oscillations given in Eq.~(\ref{eq:disprel}).
After the integration one has the following expression for $F_{nn'}$:
\begin{align}\label{eq:Fnnexpl}
  F_{nn'} = &
  \frac{4K_{0}^{2}n_{n}^{(0)}m_{i}}{3\pi^{2}\hbar\omega_{i}m_{n}}
  \left(
    \frac{m_{i}\omega_{i}}{\hbar}
  \right)^{3/2}
  \notag
  \\
  & \times
  \frac{\tilde{n}^{3/2}}{\sqrt{nn'}}
  \left(
    1+\frac{3\hbar\omega_{i}}{16m_{i}c_{s}^{2}\tilde{n}}
  \right),
\end{align}
where
$c_{s}=\lambda_{e}\omega_{i}$
is the sound
velocity and $\tilde{n}=\min(n,n')$.

Note that $F_{nn'}$ in Eq.~(\ref{eq:Fnnexpl}) is positive. It means
that the effective interaction described by the Hamiltonian in Eq.~(\ref{eq:Htotaps})
is attractive.

Now let us discuss the ions motion which corresponds to short waves.
In this situation ion-acoustic waves are transformed into Langmuir
oscillations of ions. As we have seen in Sec.~\ref{sec:QUANTSTAT},
it happens in the limit $\lambda_{e}\gg\sqrt{\hbar/m_{i}\omega_{i}}$,
that is equivalent to $c_{s}\gg\sqrt{\hbar\omega_{i}/m_{i}}$. Thus,
using Eq.~(\ref{eq:Fnnexpl}), we get that the matrix element of
the effective interaction for Langmuir oscillations takes the form,
\begin{align}\label{eq:FnnLang}
  F_{nn'}^{(\mathrm{Lang})} = &
  \frac{4K_{0}^{2}n_{n}^{(0)}m_{i}}{3\pi^{2}\hbar\omega_{i}m_{n}}
  \notag
  \\
  & \times  
  \left(
    \frac{m_{i}\omega_{i}}{\hbar}
  \right)^{3/2}
  \frac{\tilde{n}^{3/2}}{\sqrt{nn'}}.
\end{align}
Note that Eq.~(\ref{eq:FnnLang}) corrects the result of~\citet{Dvo13}. The matrix element derived by~\citet{Dvo13} was erroneous
since the incorrect dispersion relation for virtual acoustic waves
was used in the calculation of the integral in Eq.~(\ref{eq:amplint}).

To complete the analysis of the effective interaction we should define
the constant $K_{0}$. We recall that we
use the approximation of the contact interaction between ions and
neutral particles, which implies that the potential of this interaction is $K(\mathbf{r}) = K_0 \delta^3(\mathbf{r})$. Using the Born approximation~\citep{CohDiuLal77} for the scattering of ions in this potential we get the total cross section as $\sigma_{s} = K_{0}^{2} m_i^2/\pi \hbar^4$. Therefore we obtain that $K_{0}= \hbar^{2} \sqrt{\pi \sigma_{s}} /m_{i}$. The value of $\sigma_{s}$ can be extracted from experimental results in any particular case.

\section{Pairing of ions\label{sec:APPL}}

In this section we study the possibility for the formation of bound
states of ions owing to the effective attraction described in Sec.~\ref{sec:EFFINT}.
Using this effective interaction, we discuss the pairing of protons
in a dense matter of NS.

As in Sec.~\ref{sec:EFFINT}, we shall consider two ions occupying
the same energy level. On the basis of Eq.~(\ref{eq:Htotaps}) one
can see that these ions can form a bound state if the amplitude of
the effective interaction exceeds the kinetic energy of an ion,
\begin{equation}\label{eq:bscond}
  E_{n}<F_{nn},
\end{equation}
where $E_{n}$ is given in Eq.~(\ref{eq:enspectsphsym}) and $F_{nn}$
in Eq.~(\ref{eq:Fnnexpl}). Introducing the properly defined Bogoliubov
transformation, one can show that, if the condition in Eq.~(\ref{eq:bscond})
is satisfied, the ground state of the system is transformed into the
new one corresponding to the lower energy~\citep{Dvo13}.

Let us first examine the case of short ion-acoustic waves which is
equivalent to Langmuir oscillations of ions. Using Eq.~(\ref{eq:FnnLang}),
one gets that the pairing of ions takes place, i.e. Eq.~(\ref{eq:bscond})
is fulfilled, when
\begin{equation}\label{eq:n0Lang}
  n<n_{0} = \frac{4}{9\pi^{2}}\frac{\hbar m_{i}}{\omega_{i}}
  \left(
    \frac{n_{n}^{(0)}\sigma_{s}}{m_{n}}
  \right)^{2} -
  \frac{3}{2}.
\end{equation}
Thus there is an upper limit on the number of occupied states.

In Sec.~\ref{sec:QUANTSTAT} we defined the Fermi number corresponding
to the maximal possible number of occupied states. If one studies
short ion-acoustic waves, we can neglect the spatial dispersion. Thus
we can define ions wave functions in the coordinate space rather than
in the momentum space; cf. Eq.~(\ref{eq:wfasympt}). Therefore the
Fermi number can be related to the effective radius of a plasmoid
$R_{\mathrm{eff}}$ as $n_{\mathrm{F}}=m_{i}\omega_{i}R_{\mathrm{eff}}^{2}/4\hbar$~\citep{Dvo13}. If we assume that all ions inside
a plasmoid formed bound states, i.e. $n_{\mathrm{F}} < n_{0}$, we get that
\begin{equation}\label{eq:R0}
  R_{\mathrm{eff}} < R_{0} =
  \frac{4\hbar n_{n}^{(0)}\sigma_{s}}{3\pi m_{n}\omega_{i}},
\end{equation}
which is the upper bound on the plasmoid radius.

We can expect that the described phenomenon of the ions pairing can
happen in dense matter. Let us consider a spherical plasma structure
excited in the outer core of NS. It should be noted that such a background
matter is mainly composed of neutrons and has $n_{n}^{(0)}=10^{38}\thinspace\text{cm}^{-3}$.
Nevertheless various equations of state of NS matter
predict that a certain fraction of protons can be also present in
the NS core~\citep{HaePotYak12}. We shall assume that the proton density $n_{p}^{(0)}=10^{36}\thinspace\text{cm}^{-3}$,
i.e. it is about 1\% of the neutron density.

\citet{YakLevShi99} mentioned that neutrons and protons form strongly nonideal Fermi liquid in the NS outer core. We can use the formalism of quasiparticles with the effective masses $m_{i,n}^*$ for the description of plasma oscillations. \citet{OnsPea02} found that $m_{i,n}^* \approx 0.8 m_{i,n}$. Thus, using Eq.~\eqref{eq:Langfrdegpl}, we get that
$\omega_{i} \approx 1.48\times 10^{21}\thinspace\text{s}^{-1}$ in the NS outer core.

We shall be interested in the pairing of protons. The Fermi momentum of charged particles in the NS core corresponding to $n_{p}^{(0)}=10^{36}\thinspace\text{cm}^{-3}$ is $\sim 60\thinspace\text{MeV}/c$. The temperature of these particles does not exceed $10^{10}\thinspace\text{K} \sim 1\thinspace\text{MeV}$~\citep{HaePotYak12}. Thus these particles are degenerate, with protons being nonrelativistic whereas electrons being ultrarelativistic. It means that we can use our results for the description of quantum plasmoids in NS matter. Note that the application of the quantum theory is essential because of the degeneracy of the proton component of plasma.

The Fermi energy of protons in the NS outer core is $(3-6)\thinspace\text{MeV}$~\citep{HaePotYak12}. \citet{Cha06} obtained that in the MeV energy range the cross
section of the proton-neutron scattering is approximately constant
and equals to $\sigma_{s}=2\times10^{-23}\thinspace\text{cm}^{2}$. It should be noted that the scattering off a delta-function potential also gives the energy independent cross section. Thus the approximation of the contact interaction adopted in Sec.~\ref{sec:EFFINT}
is valid. Using Eq.~(\ref{eq:R0}) and accounting for the effective mass of a neutron, one obtains that $R_{0}\approx 4.47\times10^{-10}\thinspace\text{cm}$.
Note that there are $N=\tfrac{4}{3}\pi R_{0}^{3}n_{p}^{(0)}\approx 3.77\times10^{8}$
protons inside such a plasmoid. Thus the approximation of $n\gg1$,
used in our work, is valid.

Our estimate means that protons with oppositely directed spins can
form bound states. This phenomenon is similar to the formation of
Cooper pairs in metals. This result is in agreement with the predictions of various theoretical models,
recently confirmed by astronomical observations~\citep{HeiHo10},
that a superfluid and superconducting phases of nucleons can be present in the NS outer core. The conventional approach for the description of the proton superconductivity in the NS matter, reviewed, e.g., by~\citet{YakLevShi99}, implies the formation of singlet states of protons (Cooper pairs) owing to the directs nuclear interactions between these particles. \citet{YakLevShi99} mentioned that this superconducting phase can be important for the NS cooling. We suggest another mechanism for the protons pairing inside NS based on the exchange of a virtual acoustic wave. Although Cooper pairing due to direct nuclear interactions is likely to be dominant, our mechanism is not ruled out by the present knowledge about the NS structure.

Now we study the pairing of ions participating in longer ion-acoustic
waves. In this situation one should account for the effects of the
spatial dispersion. In the quantum description it is equivalent to
the consideration of the exact energy spectrum in Eq.~(\ref{eq:enspectsphsym}).
For long ion-acoustic waves, we have that $\omega<\omega_{i}$. Thus
we can expect that the effective attraction will be stronger in this
case. If we consider the case of slightly decreased frequency of
ions oscillations, i.e. $\omega\lesssim\omega_{i}$, we can still
assume that a plasmoid is localized in coordinate space and study
the corrections to $n_{0}$ and $R_{0}$ defined above. We expect
that the new values of $n_{0}$ and $R_{0}$ will increase if we account
for the spatial dispersion of ion-acoustic waves.

\begin{figure*}
    \centering
  \subfigure[]
  {\label{1a}
  \includegraphics[scale=1.1]{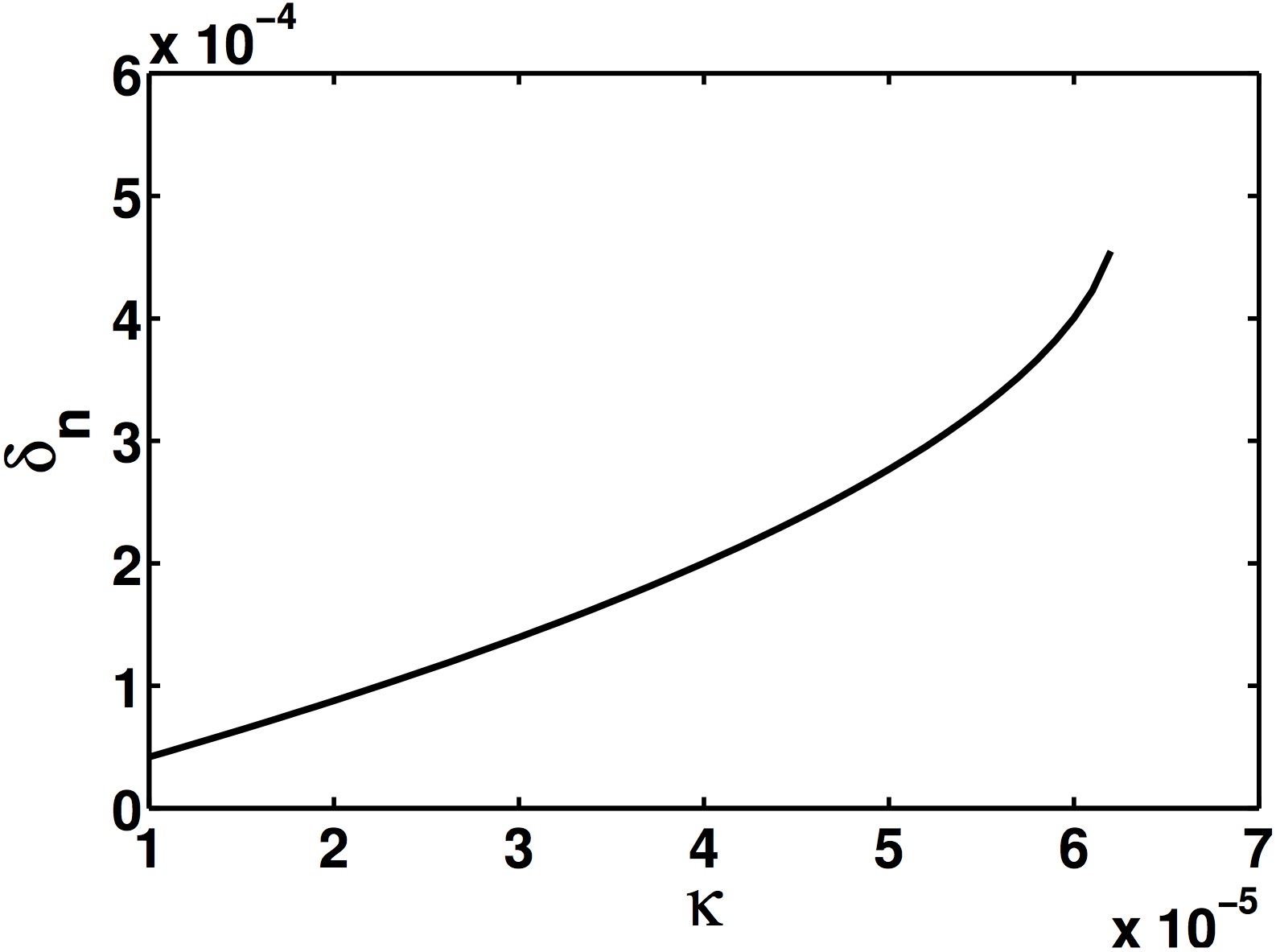}}
  \subfigure[]
  {\label{1b}
  \includegraphics[scale=1.121]{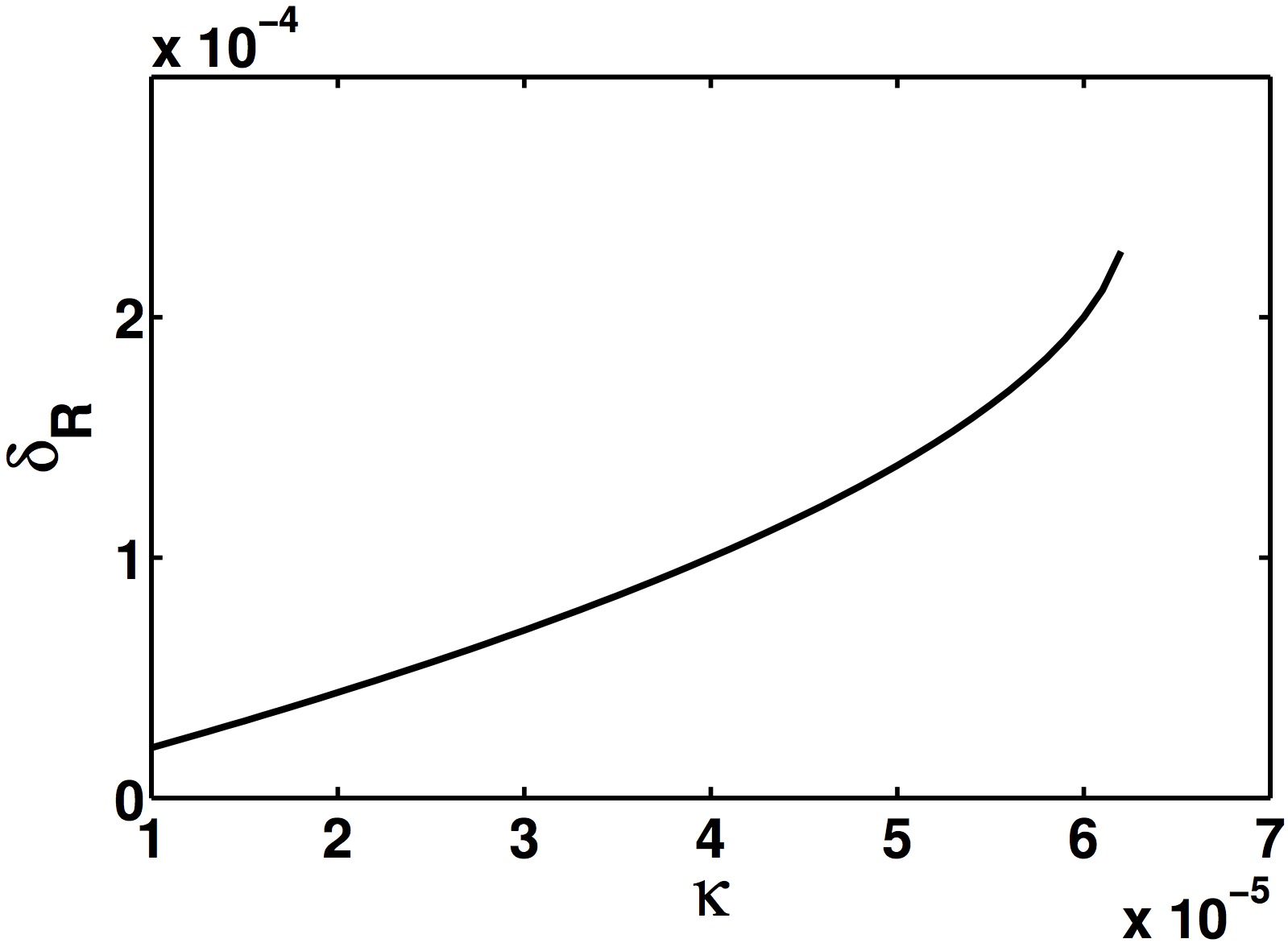}}
  \caption{The dependence of the plasmoid parameters versus $\kappa=\hbar/m_{i}\omega_{i}\lambda_{e}^{2}$
  for $n_{0}\sim10^{3}$. (a) The relative change of the critical occupation
  number $\delta_{n}=\left(n_{0}^{(\kappa)}-n_{0}\right)/n_{0}$. (b)
  The relative change of the critical plasmoid radius $\delta_{R}=\left(R_{0}^{(\kappa)}-R_{0}\right)/R_{0}$.
  \label{fig:corrn0R0}}
\end{figure*}

If the ratio $\kappa=\hbar/m_{i}\omega_{i}\lambda_{e}^{2}$ is small
but nonzero, the value of $n_{0}^{(\kappa)}$ is now determined by
the nonlinear algebraic equation $F_{nn}=E_{n}$. After we find $n_{0}^{(\kappa)}$,
we can calculate $R_{0}^{(\kappa)}$ taking that $n_{\mathrm{F}}=n_{0}^{(\kappa)}$.
In Fig.~\ref{fig:corrn0R0} we show the parameters $\delta_{n}=\left(n_{0}^{(\kappa)}-n_{0}\right)/n_{0}$
and $\delta_{R}=\left(R_{0}^{(\kappa)}-R_{0}\right)/R_{0}$, where
$n_{0}^{(\kappa)}$ and $R_{0}^{(\kappa)}$ are the critical occupation
number and the plasmoid radius corresponding to nonzero $\kappa$,
versus $\kappa$ for $n_{0}\sim10^{3}$. As one can see, if we account
for the spatial dispersion of ion-acoustic waves, the corrections
to $n_{0}$ and $R_{0}$ (we recall that $n_{0}$ and $R_{0}$ are
defined in Eqs.~(\ref{eq:n0Lang}) and~(\ref{eq:R0}) and correspond
to Langmuir oscillations of ions) are small but positive.

We have found that, for $n\gg1$, there is a very small enhancement
of the plasmoid radius if we study long ion-acoustic waves. We obtained
such a small effect because we used the ions wave functions in Eq.~(\ref{eq:wfasympt}),
which correspond to Langmuir oscillations, for the calculation of
the matrix elements. Although we have shown that these wave functions
are the correct asymptotics for $\psi_{n}$ at $n\gg1$, if we consider
great but limited values of $n$, we should use the exact $\psi_{n\sigma}(p)$
given in Eq.~(\ref{eq:wfsphsym}). In this situation the effect of
the nonzero spatial dispersion will be more significant. This fact also results from the quantum mechanical uncertainty principle. Unfortunately,
this case can be analyzed only numerically.

Let us estimate $n_\mathrm{cr}$ in Eq.~\eqref{eq:ncr} for a degenerate plasma in NS studied above. We shall consider the case when the plasmoid radius is comparable with $\lambda_e$. Of course, $R_0$ should be less than ${\lambda}_e$ for the stochastic motion of charged particles not to destroy the plasmoid. Using the value of $R_0 \sim 10^{-10} \thinspace \text{cm}$ found above we get that $n_\mathrm{cr} \sim 10^4$, which is much less than $N \sim 10^{10}$. This estimate shows that the plasma instability mentioned in Sec.~\ref{sec:QUANTSTAT} will decrease the effective plasmoid radius. Therefore this process will counteract the enhancement of the radius due to the accounting for the dispersion relation of ion-acoustic waves.

\section{Conclusion\label{sec:CONCL}}

In conclusion we mention that in the present work we have constructed
the model of a spherical quantum plasmoid based on radial oscillations
of ions. In our analysis we have suggested that the electron component
of plasma is uniformly distributed in space and ions participate in
ion-acoustic oscillations.

In Sec.~\ref{sec:QUANTSTAT} we have found the exact solution of
the Schr\"{o}dinger equation for an ion moving in the self-consistent
field of an ion-acoustic wave. We have shown that in the limiting
case of short waves the obtained wave functions of ions transform into
the previously found solution of the Schr\"{o}dinger equation for a charged
particle performing Langmuir oscillations. Then we have secondly quantized
our system. The creation and annihilation operators for oscillatory
states of ions have been introduced and the ground state of the system
has been constructed. This ground state corresponds to the collective oscillatory motion of ions. Note that the wave functions of the 3D harmonic oscillator are more appropriate for the ground state of the system since they exactly account for the dynamical features of the ions motion and the geometrical form of the plasma structure. We have also defined the Fermi number as the maximal
possible occupation number. Note that the quantization of the ions
motion is justified since later, in Sec.~\ref{sec:APPL}, we have considered
plasma structures in the very dense matter of the NS outer core, where quantum
effects are significant.

In Sec.~\ref{sec:EFFINT} we have studied the effective interaction
between oscillating ions by the exchange of a virtual acoustic wave.
Discussing the situation of plasmoids containing a very great number
of excited states $n\gg1$, we have shown that two ions occupying the
same energy level can attract each other.

The possibility of the formation of bound states of these ions have
been analyzed in Sec.~\ref{sec:APPL}. Considering the short waves
limit, we have derived the critical occupation number and the characteristic
plasmoid radius corresponding to a plasma structure in which all ions
are in bound states. As an application of our results we have discussed
the pairing of protons inside a plasmoid in the outer core of NS.
We have shown that the existence of such a plasma structure is quite
possible. It should be noted that, besides our approach for the description of the protons pairing in the outer core of NS, another mechanism for the formation of singlet states of protons, based on their direct nuclear interaction, is also discussed by~\citet{HaePotYak12}.

Since the spins of ions, which formed a bound state, should be antiparallel,
this bound state is analogous to a Cooper pair of electrons in a metal.
It is known that the formation of Cooper pairs underlies the phenomenon
of superconductivity. Our result that protons can form bound states
in the NS outer core agrees with the hypothesis that the proton superconductivity
should be present in this astrophysical environment~\citep{HaePotYak12,HeiHo10,YakLevShi99}.

It should be noted that in our work we have adopted rather simplified
analysis of the ion's motion in plasma. \citet{GalSag73} elaborated a more detailed description
of the interaction between a charged test particle and a plasma wave in
frames of the classical electrodynamics.
Various instabilities, which arise in this system, as well as the
dynamics of the turbulence were also studied by~\citet{GalSag73}
on the classical level. Our description of quantum plasmoids is valid if we neglect temperature effects, which is the case for a degenerate plasma considered in Sec.~\ref{sec:APPL}.

In Sec.~\ref{sec:APPL} we have studied the contribution of the spatial
dispersion of ion-acoustic waves to the dynamics of the system. We
have obtained that, under the assumption of great number of excited
states, this contribution to the critical occupation number and the
effective plasmoid radius is small. However, if one studies a plasma
structure with a significant but limited $n$, we expect that, e.g.,
the effective radius can be considerably enhanced.

Although we have examined plasma structures in a very dense medium
of the outer core of NS as a possible application of our results,
we may expect that the phenomenon of pairing of charged particles
can happen in a terrestrial plasma. Previously \citet{Dij80,Zel08,Dvo12} studied the formation of bound
states of charged particles to
describe some properties of stable atmospheric plasma structures.
The estimates given in Sec.~\ref{sec:APPL} show that our mechanism
of pairing cannot be directly implemented inside plasmoids in a low
density atmospheric plasma since the radius of such a structure turns
out to be quite small. Nevertheless, if one discusses a plasma structure
corresponding to a big but limited $n$, there is a possibility that the described
phenomenon can take place in a terrestrial plasma as well.

\section*{Acknowledgments}
I am thankful to the participants of the Theory Department seminar
in IZMIRAN for valuable comments, to FAPESP (Brazil) for the Grant No.~2011/50309-2,
to the Tomsk State University Competitiveness Improvement Program and to RFBR (research project No.~15-02-00293) for partial support, as well as to Y.~Kivshar for the hospitality at the ANU where this
work was partly made.

\appendix

\section{Wave functions in coordinate representation\label{sec:WFCOORD}}

In this Appendix we present the mathematical details required to
express the ion's wave function in the coordinate representation.

First we show that at $l'=0$ the wave function coincides up to a
sign factor with that found by~\citet{Dvo13}. For this purpose
we rewrite the wave function in the coordinate representation as
\begin{align}\label{eq:wfexplr}
  \psi_{n}(\mathbf{r})= & \psi_{n}(r) =
  \left[
    \frac{n!}{\Gamma(l'+n+3/2)}
  \right]^{1/2}
  \nonumber
  \\
  & \times
  \left(
    \frac{m_{i}\omega_{i}}{\hbar}
  \right)^{3/4}
  \frac{1}{\pi\beta}
  \int_{0}^{\infty}\mathrm{d}x \thinspace x^{l'+1}
  \nonumber
  \\
  & \times  
  \sin
  \left(
    \beta x
  \right)
  \exp
  \left(
    -\frac{x^{2}}{2}
  \right)
  \nonumber
  \\
  & \times  
  L_{n}^{l'+1/2}(x^{2}),
\end{align}
where we account for its spherical symmetry. Here $\beta=r/r_{0}$. Now we can obtain Eq.~(\ref{eq:wfasympt}) from Eq.~(\ref{eq:wfexplr}) at $l'=0$ using Eqs.~(8.972.3)
and~(7.388.2) on pages~1001 and~806 in~\citet{GraRyz07}.

Now let us derive the asymptotics of the wave function in Eq.~(\ref{eq:wfexplr})
in case when $0 \neq |l'|\ll1$. In this limit the integral in Eq.~(\ref{eq:wfexplr}) has the following
form:
\begin{align}\label{eq:intcs}
  \int_{0}^{\infty} & \mathrm{d}x \thinspace x^{l'+1}
  \sin
  \left(
    \beta x
  \right)
  \notag
  \\
  & \times
  \exp
  \left(
    -\frac{x^{2}}{2}
  \right)
  L_{n}^{l'+1/2}(x^{2})
  \notag
  \\
  & \approx
  \cos
  \left(
    \frac{\pi l'}{2}
  \right)
  (-1)^{n}\sqrt{\frac{\pi}{2}}\beta^{l'+1}
  \notag
  \\
  & \times
  \exp
  \left(
    -\frac{\beta^{2}}{2}
  \right)
  L_{n}^{l'+1/2}(\beta^{2})
  \notag
  \\
  & +
  \sin
  \left(
    \frac{\pi l'}{2}
  \right)J,
\end{align}
where
\begin{align}\label{eq:Hilbl}
  J = &
  \int_{0}^{\infty}\mathrm{d}x\thinspace x^{l'+1}
  \exp
  \left(
    -\frac{x^{2}}{2}
  \right)
  \notag
  \\
  & \times  
  L_{n}^{l'+1/2}(x^{2})
  \cos
  \left(
    \beta x-\frac{\pi l'}{2}
  \right).
\end{align}
To derive Eq.~(\ref{eq:intcs}) we use Eq.~(7.421.4) on page~812 in~\citet{GraRyz07}.

Note that for $|l'|\ll1$ we can set $l'=0$ in the argument of cosine
and in the upper index of the associated Laguerre polynomial in Eq.~(\ref{eq:Hilbl})
since $J$ is already multiplied by the small factor $\sin(\pi l'/2)$
in Eq.~(\ref{eq:intcs}). Thus we rewrite $J$ in the following way:
\begin{align}\label{eq:J'def}
  J \approx &
  \frac{(-1)^{n}}{2^{2n+1}n!}J',
  \notag
  \\
  J' = &
  \int_{0}^{\infty}\mathrm{d}x \thinspace
  \exp
  \left(
    -\frac{x^{2}}{2}
  \right)
  \notag
  \\
  & \times  
  H_{2n+1}(x)
  \cos
  \left(
    \beta x
  \right),
\end{align}
To study the asymptotics of $J'$ in \eqref{eq:J'def}
at $n\gg1$ we present $H_{2n+1}(x)$ in Eq.~(\ref{eq:J'def})
in the explicit form as a polynomial of $(2n+1)$th power. Then we calculate each of the integrals in the sum using Eq.~(3.952.8) on page~503 in~\citet{GraRyz07}. Finally, with help of the following expressions:
\begin{multline}\label{eq:1F1poch}
  _{1}F_{1}
  \left(
    k+1;\frac{1}{2};-\frac{\beta^{2}}{2}
  \right) 
  \\
  =
  \sum_{s=0}^{\infty}
  (-1)^{s}\frac{(k+1)_{s}}{(1/2)_{s}s!}
  \left(
    \frac{\beta^{2}}{2}
  \right)^{s},
\end{multline}
where $(z)_{s}=z(z+1)\dotsb(z+s-1)$ is the Pochhammer symbol, and
\begin{multline}\label{eq:2F1sum}
  \sum_{k=0}^{n}
  \frac{(-1)^{k}k!2^{3k}}{(2k+1)!(n-k)!}(k+1)_{s}
  \\
  =
  \frac{s!}{n!}
  {}_{2}F_{1}
  \left(
    -n,s+1;\frac{3}{2};2
  \right),
\end{multline}
where $_{2}F_{1}\left(a,b;c;z\right)$ is the Gauss hypergeometric
function,
we obtain $J'$ in the form,
\begin{align}\label{eq:J'2F1}
  J' = &
  (-1)^{n}2\sqrt{\pi}
  \frac{(2n+1)!}{n!}
  \notag
  \\
  & \times  
  \sum_{k=0}^{\infty}
  \frac{(-1)^{k}\beta^{2k}}{2^{k}\Gamma(k+1/2)}
  \notag
  \\
  & \times  
  {}_{2}F_{1}
  \left(
    -n,k+1;\frac{3}{2};2
  \right).
\end{align}
Note that one can study the limit $n\to\infty$ only in Eq.~(\ref{eq:J'2F1}).

The asymptotics at $n\gg1$ of the Gauss hypergeometric
function in Eq.~\eqref{eq:J'2F1} reads
\begin{multline}\label{eq:2F1asyptI}
  _{2}F_{1}
  \left(
    -n,k+1;\frac{3}{2};2
  \right)
  \\
  \sim
  (-1)^{n}\frac{2^{k-1}}{k!}\sqrt{\frac{\pi}{2}}
  \left(
    n+\frac{3}{2}
  \right)^{k-1/2}.
\end{multline}
To derive Eq.~\eqref{eq:2F1asyptI} we take into account that
\begin{align}\label{eq:2F1Euler}
  & _{2}F_{1}
  \left(
    a,b;c;z
  \right)
  =
  \frac{\Gamma(c)\Gamma(b-a)}{\Gamma(b)\Gamma(c-a)}
  (-1)^{a}z^{-a}
  \notag
  \\
  & \times  
  {}_{2}F_{1}
  \left(
    a,a+1-c;a+1-b;\frac{1}{z}
  \right)
  \nonumber
  \\
  & +
  \frac{\Gamma(c)\Gamma(a-b)}{\Gamma(a)\Gamma(c-b)}
  (-1)^{b}z^{-b}
  \notag
  \\
  & \times  
  {}_{2}F_{1}
  \left(
    b,b+1-c;b+1-a;\frac{1}{z}
  \right),
\end{align}
and
\begin{align}
  \frac{\Gamma(-n-k-1)}{\Gamma(-n)} 
  = &
  (-1)^{k+1}
  \notag
  \\
  & \times
  \frac{n!}{(n+k+1)!}.
\end{align}
We also use the following asymptotics:
\begin{align}
  _{2}F_{1} &
  \left(
    -n,-n-\frac{1}{2};-n-k;\frac{1}{2}
  \right)
  \notag
  \\
  & \sim
  2^{-n+k-1/2},
  \nonumber
  \\
  _{2}F_{1} &
  \left(
    k+1,k+\frac{1}{2};n+k+2;\frac{1}{2}
  \right)
  \notag
  \\
  & \to 1,
\end{align}
which are valid at $n \gg 1$.

It is interesting to compare Eq.~(\ref{eq:2F1asyptI}) with Eq.~(2.7)
derived by~\citet{Tem86}, where the asymptotics of the Gauss hypergeometric
function was also studied. The result of~\citet{Tem86} reads
\begin{multline}\label{eq:2F1asyptTem}
  _{2}F_{1}
  \left(
    -n,k+1;\frac{3}{2};2
  \right)
  \\
  \sim
  (-1)^{n} \frac{2^{k-1}}{k!}\sqrt{\frac{\pi}{2}}
  \left(
    n+1
  \right)^{k-1/2}.
\end{multline}
We can see that besides the unessential difference, $n+3/2$ vs. $n+1$,
which is unimportant at big $n$, our result coincides with that of obtained by~\citet{Tem86}.
However, \citet{Tem86} claimed that the asymptotics
in question is valid when the argument of the hypergeometric function
$z>2$. Here we demonstrate that the case $z=2$ can be
also described by Eq.~(\ref{eq:2F1asyptI}) or~(\ref{eq:2F1asyptTem})
at least for the particular set of the parameters of the hypergeometric
function.

\begin{figure*}
  \includegraphics{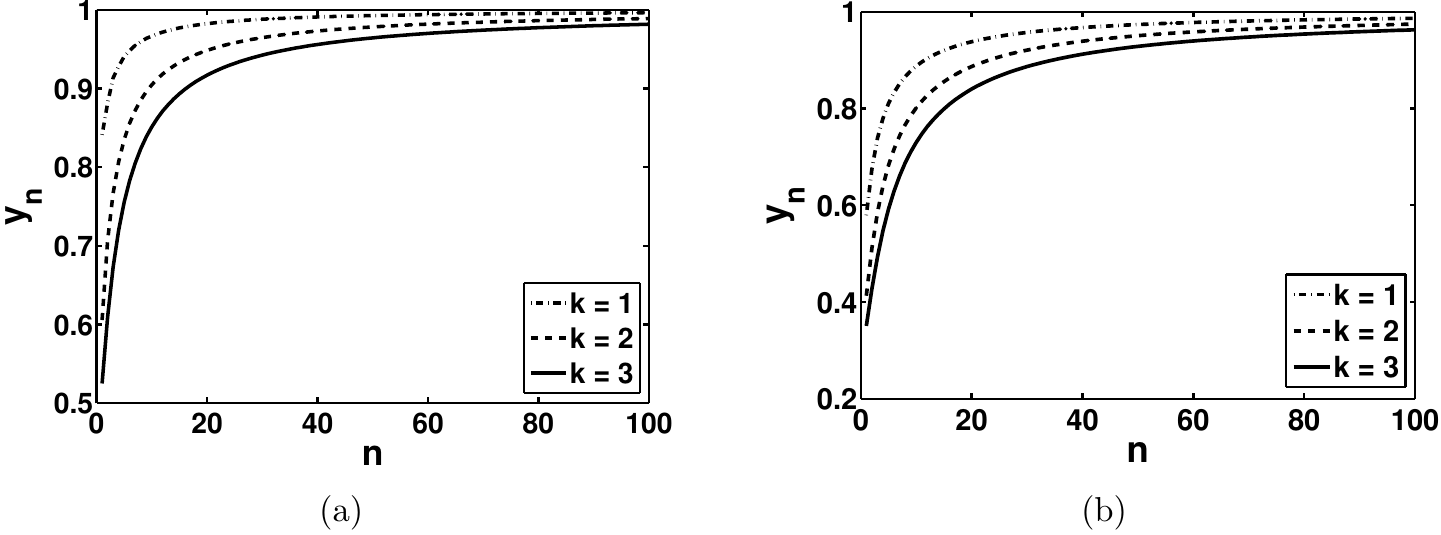}\caption{The illustration of the asymptotic behavior of the hypergeometric
  functions in Eqs.~(\ref{eq:2F1asyptI}) and~(\ref{eq:3F2asympt})
  for different values of $k$. (a) The sequence $y_{n}$ in Eq.~(\ref{eq:yn2F1}).
  (b) The sequence $y_{n}$ in Eq.~(\ref{eq:yn3F2}).\label{fig:hypergeomas}}
\end{figure*}

The asymptotic behavior of the hypergeometric function in Eq.~(\ref{eq:2F1asyptI})
is shown in Fig.~\ref{fig:hypergeomas}(a). We depict there the sequence
\begin{align}\label{eq:yn2F1}
  y_{n} = &
  (-1)^{n}\sqrt{\frac{2}{\pi}}
  \notag
  \\
  & \times  
  \frac{k!
  {}_{2}F_{1}
  \left(
    -n,k+1;3/2;2
  \right)
  }
  {2^{k-1}
  \left(
    n+3/2
  \right)^{k-1/2}
  },
\end{align}
for different values of $k$. One can see in Fig.~\ref{fig:hypergeomas}(a)
that $y_{n}\to1$. It proves the validity of the asymptotics in Eq.~(\ref{eq:2F1asyptI}).

\begin{figure*}
  \centering
  \subfigure[]
  {\label{3a}
  \includegraphics[scale=.35]{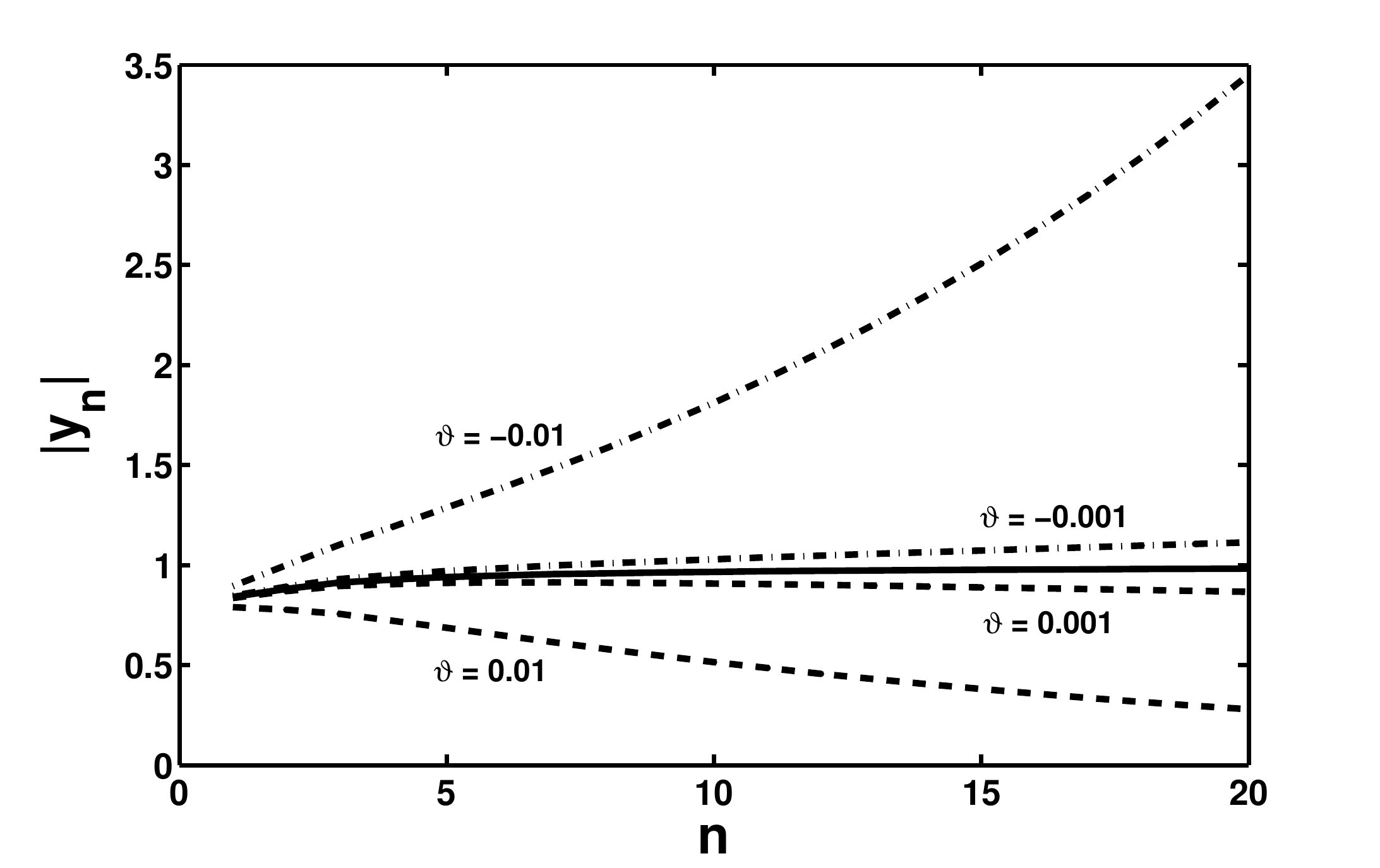}}
  \hskip-.7cm
  \subfigure[]
  {\label{3b}
  \includegraphics[scale=.35]{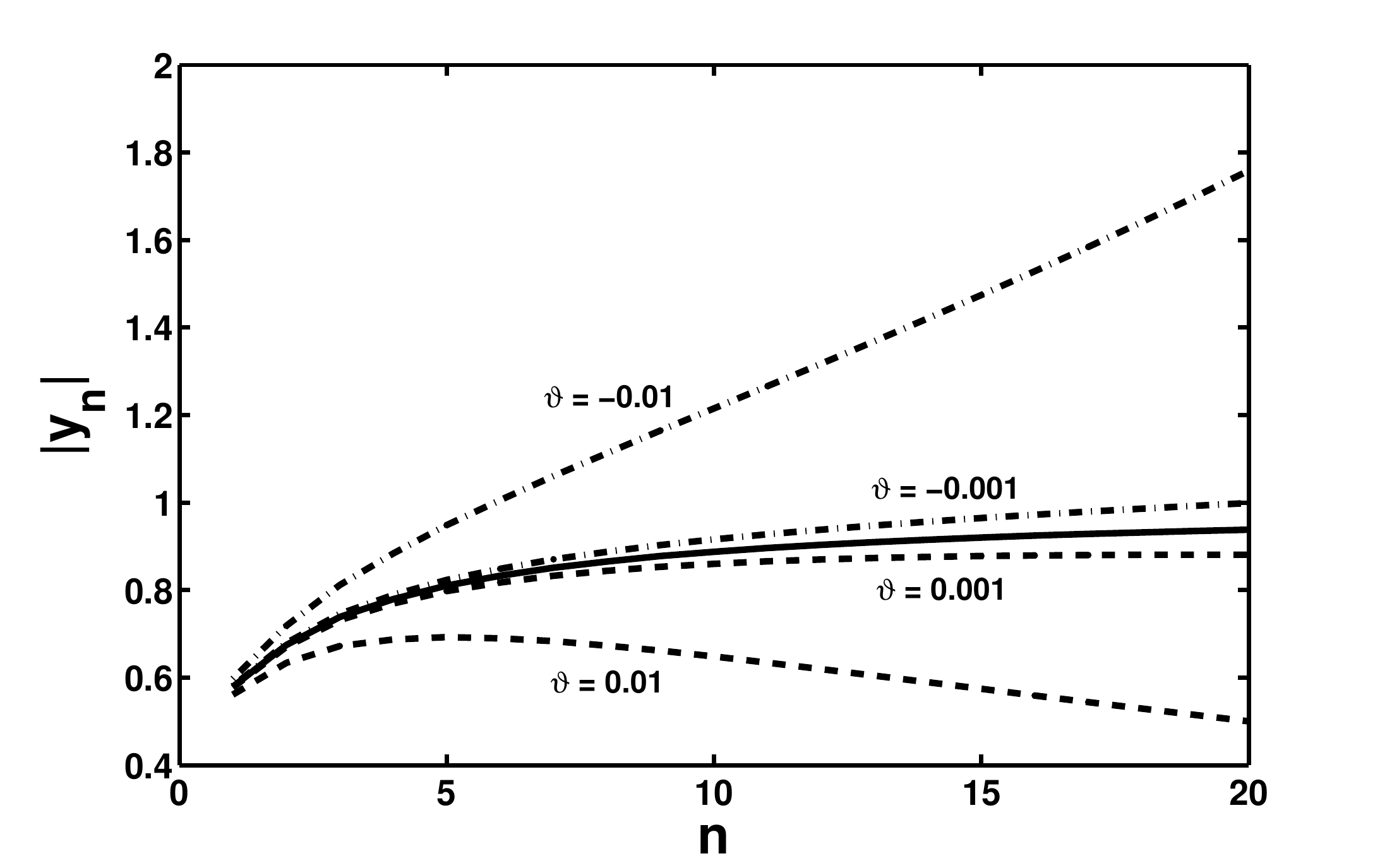}}
  \caption{(a) The behavior of the absolute value of $y_n$ in Eq.~\eqref{eq:yn2F1} with the complex radial quantum number $n \to n e^{\mathrm{i}\vartheta}$. Dashed lines correspond to $\vartheta > 0$ and dash-dotted to $\vartheta < 0$. The solid line is built for $\vartheta = 0$ and is also shown in Fig.~\ref{fig:hypergeomas}(a). (b) The same as in panel~(a) for $y_n$ in Eq.~\eqref{eq:yn3F2}.
  \label{yncomplex}}
\end{figure*}

To illustrate the behavior of $y_n$ in Eq.~\eqref{eq:yn2F1} at big $n\gg 1$ we can formally take that the radial quantum number is a complex number $n \to n e^{\mathrm{i}\vartheta}$, where $\vartheta$ is the phase. In Fig.~\ref{yncomplex}(a) we plot $|y_n|$ versus $n$ for different values of $\vartheta$ and $k=1$. One can see that the dotted and dash-dotted lines, which correspond to positive and negative $\vartheta$, approach to the solid line built for $\vartheta = 0$ at $|\vartheta| \to 0$. Therefore, the numerical experiment demonstrates that $y_n$ in Eq.~\eqref{eq:yn2F1} in a smooth function in the complex plane. Of course, a more careful analytical analysis of this fact is required. However, this issue is beyond the scope of the present work.

On the basis of Eqs.~(\ref{eq:1F1poch})-(\ref{eq:2F1asyptI}) we
obtain the behavior of $J'$ as
\begin{equation}\label{eq:J'fin}
  J' \sim \sqrt{2}2^{2n}n!
  \cos
  \left(
    2\sqrt{n+3/2}\beta
  \right),
\end{equation}
where we use the value of the sum of the series,
\begin{multline}
  \sum_{k=0}^{\infty}
  \frac{(-1)^{k}}{k!\Gamma(k+1/2)}
  \left(
    \sqrt{n+3/2}\beta
  \right)^{2k}
  \\
  =
  \frac{1}{\sqrt{\pi}}\cos(2\sqrt{n+3/2}\beta).
\end{multline}
Using Eqs.~(\ref{eq:intcs}), (\ref{eq:J'def}), and~(\ref{eq:J'fin}),
one gets the asymptotics of the wave function in Eq.~(\ref{eq:wfexplr}),
which corresponds to the case $|l'|\ll1$ and $n\gg1$,
\begin{align}\label{eq:wfasyptl0}
  \psi_{n}(r) \sim &
  \frac{(-1)^{n}}{\sqrt{2}\pi\beta n^{1/4}}
  \left(
    \frac{m_{i}\omega_{i}}{\hbar}
  \right)^{3/4}
  \notag
  \\
  & \times
  \sin
  \left(
    2\sqrt{n}\beta
  \right),
\end{align}
where we keep only the leading term in $l'$ and use Eq.~(8.978.3)
on page~1003 in~\citet{GraRyz07}. One can see in Eq.~(\ref{eq:wfasyptl0}) that the correction for
the wave function, linear in $l'$, vanishes in the limit $n\gg1$.

Now let us consider another extreme situation which corresponds to
$l'=-1/2$. On the basis of Eq.~(\ref{eq:wfexplr}) one can see that
in this case it is necessary to get the asymptotics at big $n$ of
the following integral:
\begin{align}\label{eq:Jl1/2}
  J = &
  \int_{0}^{\infty}\mathrm{d}x \thinspace
  \sqrt{x}
  \sin
  \left(
    \beta x
  \right)
  \notag
  \\
  & \times  
  \exp
  \left(
    -\frac{x^{2}}{2}
  \right)
  L_{n}(x^{2}),
\end{align}
where $L_{n}(z)=L_{n}^{0}(z)$ is the Laguerre polynomial. We can
analyze $J$ in Eq.~(\ref{eq:Jl1/2}) in the same manner as $J'$
in Eq.~(\ref{eq:J'def}). We shall describe only the main steps of
this analysis.

First, we represent $L_n(x^2)$ in the explicit form as a polynomial of $2n$th power.
Then we use Eq.~(3.952.7) on page~503 in~\citet{GraRyz07} and the sum of the series,
\begin{align}\label{eq:ser3F2}
  \sum_{k=0}^{n} &
  \frac{(-1)^{k}2^{k}}{k!^{2}(n-k)!}
  \Gamma
  \left(
    k + \frac{5}{4}
  \right)
  \left(
    k+\frac{9}{4}
  \right)_{s}
  \notag
  \\
  & =
  \frac{4}{5}
  \frac{\Gamma
  \left(
    s+\frac{9}{4}
  \right)
  }
  {n!}
  \notag
  \\
  & \times
  {}_{3}F_{2}
  \left(
    -n,s+\frac{9}{4},\frac{5}{4};1,\frac{9}{4};2
  \right),
\end{align}
where $_{3}F_{2}\left(a_{1},a_{2},a_{3};b_{1},b_{2};z\right)$ is
the generalized hypergeometric function.
Eventually we get the expression for $J$ in the form,
\begin{align}\label{eq:J3F2}
  J = &
  \frac{2^{5/4}}{5}\sqrt{\pi}\beta
  \sum_{k=0}^{\infty}
  \frac{(-1)^{k}\beta^{2k}\Gamma(k+9/4)}{2^{k}k!\Gamma(k+3/2)}
  \notag
  \\
  & \times  
  {}_{3}F_{2}
  \left(
    -n,k+\frac{9}{4},\frac{5}{4};1,\frac{9}{4};2
  \right).
\end{align}
Here we also use the analog of Eq.~(\ref{eq:1F1poch}).

The hypergeometric function in Eq.~(\ref{eq:J3F2}) has the following
behavior at $n\gg1$:
\begin{align}\label{eq:3F2asympt}
  _{3}F_{2} &
  \left(
    -n,k+\frac{9}{4},\frac{5}{4};1,\frac{9}{4};2
  \right)
    \notag
  \\
  & \sim
  5
  \left(
    n+1
  \right)^{1/4}
  \left(
    n+\frac{9}{4}
  \right)^{k}
  \notag
  \\
  & \times  
  \frac{(-1)^{n}2^{k-7/4}}{\Gamma(k+9/4)}.
\end{align}
To illustrate the asymptotics of the hypergeometric function in Eq.~(\ref{eq:3F2asympt}),
in Fig.~\ref{fig:hypergeomas}(b) we present the sequence
\begin{align}\label{eq:yn3F2}
  y_{n} = &
  (-1)^{n}
  \frac{\Gamma(k+9/4)}
  { 5 \times 2^{k-7/4}}
  \\
  \notag
  & \times
  \frac{
  {}_{3}F_{2}
  \left(
    -n,k+9/4,5/4;1,9/4;2
  \right)
  }
  {
  (n+1)^{1/4}
  \left(
    n+9/4
  \right)^{k}},
\end{align}
for different values of $k$. One can see in Fig.~\ref{fig:hypergeomas}(b)
that $y_{n}\to1$, as it follows from Eq.~(\ref{eq:3F2asympt}).

Analogously to Fig.~\ref{yncomplex}(a), in Fig.~\ref{yncomplex}(b) we show the behavior of $|y_n|$ in Eq.~\eqref{eq:yn3F2} for the complex valued radial quantum number $n \to n e^{\mathrm{i}\vartheta}$ at $k=1$. Again one can see that $|y_n| \to |y_n(\vartheta = 0)|$ at $|\vartheta| \to 0$, i.e. $y_n$ is a smooth function in the complex plane.

Finally, using Eqs.~(\ref{eq:Jl1/2})-(\ref{eq:3F2asympt}) one
obtains the expression for the wave function at $n\gg1$,
\begin{align}\label{eq:wfasymptl1/2}
  \psi_{n}(r) \sim &
  \frac{(-1)^{n}}{\sqrt{2}\pi\beta n^{1/4}}
  \left(
    \frac{m_{i}\omega_{i}}{\hbar}
  \right)^{3/4}
  \notag  
  \\
  & \times
  \sin
  \left(
    2\sqrt{n}\beta
  \right).
\end{align}
To derive Eq.~(\ref{eq:wfasymptl1/2}) we use the known value for
the sum of the series,
\begin{multline}
  \sum_{k=0}^{\infty}
  \frac{(-1)^{k}}{k!\Gamma(k+3/2)}
  \left(
    \sqrt{n+9/4}\beta
  \right)^{2k}
  \\
  =
  \frac{\sin(2\sqrt{n+9/4}\beta)}{\sqrt{\pi}\beta\sqrt{n+9/4}}.
\end{multline}
One can see in Eq.~(\ref{eq:wfasymptl1/2}) that, in the limit $n\gg1$,
the expression for the wave function corresponding to $l'=-1/2$
again coincides with the analogous expression for $l'=0$.


At the end of this section we mention that as a by-product have obtained
the asymptotic expression for the Hilbert transform of the Hermite
function with the odd index,
\begin{align}\label{eq:Hermfun}
  \varphi_{2n+1}(z) = &
  \frac{1}{\pi^{1/4}\sqrt{2^{n}n!}}
  \notag
  \\
  & \times
  \exp
  \left(
    -\frac{z^{2}}{2}
  \right)
  H_{2n+1}(z).
\end{align}
The Hilbert transform $H[f](\beta)$ of the function $f(x)$ is defined
as
\begin{equation}\label{eq:Hilbtrans}
  H[f](\beta) =
  \frac{1}{\pi}
  \mathrm{V.P.}
  \int_{-\infty}^{+\infty} \mathrm{d}x
  \frac{f(x)}{\beta-x},
\end{equation}
where $\mathrm{V.P.}$ stays for the principle value of the integral.

Using Eqs.~(\ref{eq:J'fin})
and~(\ref{eq:Hermfun})
we obtain the following asymptotics of the odd index Hilbert transform of the
Hermite function:
\begin{align}\label{eq:HilbHermasympt}
  H[\varphi_{2n+1}](\beta) \sim &
  (-1)^{n+1}\frac{2^{3n/2+1}\sqrt{n!}}{\pi^{3/4}}
  \notag
  \\
  & \times  
  \cos
  \left(
    2\sqrt{n+3/2}\beta
  \right).
\end{align}
It should be noted that previously only the recurrence relation for
$H[\varphi_{2n+1}]$ was know~\citep{Hah00}, which
can be used only for small $n$. We have derived the explicit expression
for the asymptotics of the Hilbert transform. Therefore, Eq.~(\ref{eq:HilbHermasympt})
can be useful, for instance, in signal processing.

\balance


\begin{thebibliography}{100}

\bibitem[Alhaidari(2002)]{Alh02}
  Alhaidari~A.~D. 2002
  Solutions of the nonrelativistic wave equation with position-dependent effective mass
  \textit{Phys. Rev.} A \textbf{66} 042116.

\bibitem[Baiko(2009)]{Bai09}
  Baiko~D.~A. 2009
  Coulomb crystals in the magnetic field
  \textit{Phys. Rev.} E \textbf{80} 046405 
  [arXiv:0910.0171].

\bibitem[Birkl \textit{et al.}(1992)]{BirKasWal92}
  Birkl~G., Kassner~S. \& Walther~H. 1992
  Multiple-shell structures of laser-cooled $^{24}$Mg$^{+{}}$
  ions in a quadrupole storage ring
  \textit{Nature} \textbf{357} 310--313.

\bibitem[Blokhintsev(1964)]{Blo64}
  Blokhintsev,~D.~I. 1964
  \textit{Quantum Mechanics}
  (Dordrecht: Reidel) pp.~140--1.

\bibitem[Bonitz \textit{et al.}(2008)]{Bon08}
  Bonitz,~M., \textit{et al.} 2008
  Classical and quantum Coulomb crystals
  \textit{Phys. Plasmas} \textbf{15} 055704
  [arXiv:0801.0754].

\bibitem[Bonitz \textit{et al.}(2013)]{BonPehSch13}
  Bonitz,~M., Pehlke,~E. \& Schoof,~T. 2013
  Attractive forces between ions in quantum plasmas:
  Failure of linearized quantum hydrodynamics
  \textit{Phys. Rev.} E \textbf{87} 033105
  [arXiv:1205.4922].

\bibitem[Braaten \& Segel(1993)]{BraSeg93}
  Braaten,~E. \& Segel,~D. 1993
  Neutrino energy loss from the plasma process at all temperatures and densities
  \textit{Phys. Rev.} D \textbf{48}, 1479--1491
  [hep-ph/9302213].

\bibitem[Carstensen \textit{et al.}(2012)]{Car12}
  Carstensen,~J., \textit{et al.} 2012
  Charging and coupling of a vertically aligned particle pair in the plasma sheath
  \textit{Phys. Plasmas} \textbf{19} 033702.

\bibitem[Chadwick \textit{et al.}(2006)]{Cha06}
  Chadwick,~M.~B., \textit{et al.} 2006
  ENDF/B-VII.0: next generation evaluated nuclear data library for nuclear science
  and technology
  \textit{Nucl. Data Sheets} \textbf{107} 2931--3059.

\bibitem[Chu \& I(1994)]{ChuL94}
  Chu,~J.~H., \& I,~L. 1994
  Direct observation of Coulomb crystals and liquids
  in strongly coupled rf dusty plasmas
  \textit{Phys. Rev. Lett.} \textbf{72} 4009--4012.

\bibitem[Cohen-Tannoudji \textit{et al.}(1977)]{CohDiuLal77}
  Cohen-Tannoudji,~C., Diu,~B. \& Lalo\"e,~F. 1977
  \textit{Quantum Mechanics} vol.~2
  (New York, NY: Wiley) pp.~919--920.

\bibitem[Dawson(1959)]{Daw59}
  Dawson,~J.~M. 1959
  Nonlinear electron oscillations in a cold plasma
  \textit{Phys. Rev.} \textbf{113} 383--387.

\bibitem[Dijkhuis(1980)]{Dij80}
  Dijkhuis,~G.~C. 1980
  A model for ball lightning
  \textit{Nature} \textbf{284} 150--151.

\bibitem[Dvornikov(2012)]{Dvo12}
  Dvornikov,~M. 2012
  Effective attraction between oscillating electrons
  in a plasmoid via acoustic wave exchange
  \textit{Proc. R. Soc.} A \textbf{468} 415--428
  [arXiv:1102.0944].

\bibitem[Dvornikov(2013)]{Dvo13}
  Dvornikov,~M. 2013
  Pairing of charged particles in a quantum plasmoid
  \textit{J. Phys. A: Math. Theor.} \textbf{46} 045501
  [arXiv:1208.2208].

\bibitem[Dvornikov \& Dvornikov(2006)]{DvoDvo06}
  Dvornikov,~M., \& Dvornikov,~S. 2006
  Electron gas oscillations in plasma: Theory and applications
  \textit{Advances in Plasma Physics Research} vol.~5, ed. F.~Gerard
  (New York, NY: Nova Science Publishers) pp~197--212
  [physics/0306157].

\bibitem[Gradshteyn \& Ryzhik(2007)]{GraRyz07}
  Gradshteyn,~I.~S. \& Ryzhik,~I.~M. 2007
  \textit{Table of Integrals, Series, and Products}
  7th edn. (Amsterdam: Elsevier).

\bibitem[Grimes \& Adams(1979)]{GriAda79}
  Grimes,~C.~C., \& Adams,~G. 1979
  Evidence for a liquid-to-crystal phase transition in a classical,
  two-dimensional sheet of electrons
  \textit{Phys. Rev. Lett.} \textbf{42} 795--798.

\bibitem[Haas(2011)]{Haa11}
  Haas,~F. 2011
  \textit{Quantum Plasmas: An Hydrodynamic Approach}
  (New York, NY: Springer).

\bibitem[Haensel \textit{et al.}(2007)]{HaePotYak12}
  Haensel,~P., Potekhin,~A.~Y. \& Yakovlev,~D.~G. 2007
  \textit{Neutron Stars I. Equation of State and Structure}
  (New York, NY: Springer) pp.~207--280.

\bibitem[Hahn(2000)]{Hah00}
  Hahn,~S.~L. 2000
  Hilbert transform
  \textit{The Transforms and Applications Handbook}
  2nd edn., ed. A.~D.~Poularikas
  (Boca Raton, FL: CRC Press) ch.~7.10.

\bibitem[Heinke \& Ho(2010)]{HeiHo10}
  Heinke,~C.~O. \& Ho,~W.~C.~G. 2010
  Direct observation of the cooling of the Cassiopeia A neutron star
  \textit{Astrophys. J}. \textbf{719} L167--L171
  [arXiv:1007.4719].

\bibitem[Lifshitz \& Pitaevski\u{\i}(2010)]{LifPit10}
  Lifshitz,~E.~M. \& Pitaevski\u{\i},~L.~P. 2010
  \textit{Physical Kinetics}
  (Burlington, MA: Elsevier) pp.~136--137.

\bibitem[Morfill \& Ivlev(2009)]{MorIvl09}
  Morfill,~G.~E. \& Ivlev,~A.~V. 2009
  Complex plasmas: An interdisciplinary research field
  \textit{Rev. Mod. Phys.} \textbf{81} 1353--1404.

\bibitem[Nambu \textit{et al.}(1995)]{NamVlaShu95}
  Nambu,~M., Vladimirov,~S.~V. \& Shukla,~P.~K. 1995
  Attractive forces between charged particulates in plasmas
  \textit{Phys. Lett.} A \textbf{203} 40--42.



\bibitem[Onsi \& Pearson(2002)]{OnsPea02}
  Onsi,~M. \& Pearson,~J.~M. 2002
  Equation of state of stellar nuclear matter and the effective nucleon mass
  \textit{Phys. Rev.} C \textbf{65} 047302.

\bibitem[Sagdeev \& Galeev(1969)]{GalSag73}
  Sagdeev,~R.~Z. \& Galeev,~A.~A. 1969
  \textit{Nonlinear Plasma Theory}
  (New York, NY: W.~A.~Benjamin, Inc.) pp.~37--113.


\bibitem[Schmidt(2007)]{Sch07}
  Schmidt,~A.~G.~M. 2007
  Time evolution for harmonic oscillators with position-dependent mass
  \textit{Phys. Scr.} \textbf{75} 480--483.

\bibitem[Temme(1986)]{Tem86}
  Temme,~N.~M. 1986
  Uniform asymptotic expansion for a class of polynomials biorthogonal
  on the unit circle
  \textit{Constr. Approx.} \textbf{2} 369--376.

\bibitem[Vlasov \& Yakovlev(1978)]{VlaYak78}
  Vlasov,~A.~A. \& Yakovlev,~M.~A. 1978
  Interaction between ions through an intermediate system (neutral gas)
  and problem of existence of a cluster of particles maintained
  by its own forces: I
  \textit{Theor. Math. Phys.} \textbf{34} 124--130.

\bibitem[Yakovlev \textit{et al.}(1999)]{YakLevShi99}
  Yakovlev,~D.~G., Levenfish,~K.~P. \& Shibanov,~Yu.~A. 1999
  Cooling neutron stars and superfluidity in their interiors
  Phys.--Usp. \textbf{42} 737--778
  [astro-ph/9906456].

\bibitem[Zelikin(2008)]{Zel08}
  Zelikin,~M.~I. 2008
  Superconductivity of plasma and fireballs
  \textit{J. Math. Sci.} \textbf{151} 3473--3496.

\end{thebibliography}
\end{document}